\documentclass[referee]{raa}            
\usepackage[dvips]{graphicx}
\usepackage{multirow}
\usepackage{dcolumn}
\usepackage{natbib}
\bibpunct{(}{)}{;}{a}{}{,}
\usepackage{ulem}
\usepackage{bm}
\usepackage{amsmath}
\usepackage{array}

\usepackage{graphicx,times}             

\begin{document}

   \title{Numerical Simulation of Superhalo Electrons Generated by Magnetic Reconnection in the Solar Wind Source Region
\footnotetext{$*$ Supported by NSFC and PKU.}
}
   \volnopage{Vol.0 (200x) No.0, 000--000}      
   \setcounter{page}{1}          

   \author{L.-P. Yang
      \inst{1,2}
   \and L.-H. Wang
      \inst{1}
   \and J.-S. He
      \inst{1}
   \and C.-Y. Tu
      \inst{1}
   \and S.-H. Zhang
      \inst{3}
   \and L. Zhang
      \inst{1}
   \and X.-S. Feng
      \inst{2}
   }

   \institute{School of Earth and Space Sciences, Peking University, 
100871 Beijing, China; {\it wanglhwang@gmail.com}\\
        \and
             SIGMA Weather Group, State Key Laboratory for Space
Weather, Center for Space Science and Applied Research, Chinese
Academy of Sciences, 100190 Beijing, China\\
        \and
             Institute of Geology and Geophysics, Chinese
Academy of Sciences, 100029 Beijing, China\\
   }

   \date{Received~~; accepted~~}

\abstract{ Superhalo electrons appear to be continuously present
in the interplanetary medium, even at very quiet times, with a power-law spectrum at energies
above $\sim$2 keV. Here we numerically investigate the generation of superhalo electrons by magnetic reconnection in the solar wind source region, using the MHD and test particle simulations for both single X-line reconnection and multiple X-line reconnection. We
find that the direct current electric field, produced in the
magnetic reconnection region, can accelerate electrons from an
initial thermal energy of T $\sim10^5$ K up to hundreds of keV.
After acceleration, some of the accelerated electrons, together with
the nascent solar wind flow driven by the reconnection, propagate
upwards along the newly-opened magnetic field lines into the
interplanetary space, while the rest move downwards into the lower
atmosphere. Similar to the observed superhalo electrons at 1 AU, the
flux of the upward-traveling accelerated electrons versus energy
displays a power-law distribution at $\sim$ 2 $-$ 100 keV, $f(E)
\sim E^{-\delta}$, with a $\delta$ of $\sim$ 1.5 $-$ 2.4. For single
(multiple) X-line reconnection, the spectrum becomes harder (softer)
as the anomalous resistivity parameter $\alpha$ (uniform resistivity
$\eta$) increases. These modeling results suggest that the acceleration in the solar
wind source region may contribute to superhalo electrons.
}

   \authorrunning{L.-P. Yang et al. }            
   \titlerunning{Numerical Simulation of Superhalo Electrons in the Solar Wind Source Region }  

   \maketitle

%
%
\section{Introduction}           
\label{sect:intro}

Electron measurements from the 3D Plasma and Energetic Particle instrument on the WIND spacecraft near 1 AU find a superthermal
component of solar wind electron population at energies above
$\sim$2 keV, denoted the ``superhalo'', with a power-law ($f(E) \sim
E^{-\delta}$ with $\delta \sim 2.5$) spectrum extending to $>$ 100
keV, and a nearly isotropic angular distribution \citep{Lin1997,
Lin1998}. Superhalo electrons appear to be the electron counterpart of the 
power-law-tail suprathermal ions above solar wind and pickup ion energies 
that are observed throughout the heliosphere at all the times \citep[e.g.,][]{Gloeckler2008, Wimmer2013}.
Using high sensitivity measurements from the SupraThermal
Electron instrument \citep{Lin2008} on the STEREO, \cite{Wang2012}
reported that the power-law spectral index $\delta$ of superhalo
electrons observed during quiet-times near solar minimum ranges from
$\sim$ 1.5 to $\sim$ 3.4, with an average of $\sim2.35\pm0.45$. The
observed density of superhalo electrons, about $10^{-9} - 10^{-6}$
of the solar wind proton density, decreases with the decay of solar
cycle, while $\delta$ has no solar-cycle variation. Since these
superhalo electrons are present even in the absence of any solar
activity (e.g., active regions, flares, etc.), \cite{Wang2012}
suggested that superhalo electrons may be generated by wave-particle
interactions in the interplanetary medium (IPM), or by nonthermal processes related to the acceleration of the solar wind.

With a weak turbulence approach, \cite{Yoon2012} proposed that
superhalo electrons are accelerated by local resonant interactions
with electron beam-excited Langmuir waves and that the dynamic
equilibrium between these electrons and Langmuir waves predicts a
power-law spectrum of $E^{-2.3}$, consistent with observations.
However, \cite{Podesta2008} argued that the observed energy density
of the Langmuir waves at 1AU is too small to accelerate superhalo electrons by the time the solar wind reaches 1 AU.

If superhalo electrons arise from the Sun, one possible accelerator
is the magnetic reconnection in the solar wind source region. Many
researches have examined the charged-particle acceleration by
magnetic reconnection in solar flares or the Earth's magnetosphere
\citep[e.g.,][]{Speiser1965, Bulanov1976, Bulanov1980, Martens1990,
Litvinenko1993,  Miller1997, Mori1998, Browning2001, Hamilton2003,
Zharkova2004, Wood2005, Turkmani2006, Cargill2006, Onofri2006,
Liu2009, Drake2010, Gordovskyy2010, Oka2010, Kowal2011, Li2012,
Bian2013, Leonardis2013}. \cite{Speiser1965} was the first to
analyze the direct current (DC) electric field acceleration in the
single reconnecting current sheet (RCS), by analytically solving the
particle motion equations in the geomagnetic tail. Based on
simulations with the single RCS in solar flare, \cite{Mori1998}
obtained a power-law energy spectrum with the index of 2.0-2.2 for
the accelerated protons, consistent with the theoretical prediction
by \cite{Bulanov1980}; \cite{Wood2005} got a power-law energy
spectrum with the index $\sim$ 1.5 for the accelerated electrons in
the reconnection outflow region. Moreover, \cite{Turkmani2006} and
\cite{Cargill2006} found a power-law distribution for the
accelerated ions and electrons, from simulations with the stochastic
development of transient RCSs. \cite{Onofri2006} also suggested that
the fragmented RCSs can be very efficient electron accelerator.
Based on two-dimensional particle-in-cell (PIC) simulations of
multi-island reconnection, \cite{Drake2010} and \cite{Oka2010}
proposed that electron and ion acceleration is dominated by Fermi
reflection in contracting and merging magnetic islands. However, the
above models related to solar flares may not conform well to the
superhalo electrons that are not correlated with flares. Due to
their continuous presence in the PM, similar to the solar wind,
these superhalo electrons could be produced by the magnetic
reconnection related to the solar wind origin \citep{Wang2012}.

If the superhalo electrons observed at 1 AU originate from the solar
wind source region, then a self-consistent acceleration model would
also include the particle-escape process from the Sun to the IPM.
One possibility is to involve open magnetic field lines out into the
IPM, as suggested by many models of solar energetic particles from
the acceleration in transient events such as solar flares and
coronal mass ejections \citep{Vainio2000, Dmitruk2003, Arzner2006,
Rosdahl2010, Baumann2012, Masson2012}. Recently, we have simulated
the reconnection between a closed loop and an open funnel in the
solar wind source region, to examine the origin of the solar wind
\citep{Yang2013}. In the present study, we utilize this magnetic
reconnection model (Section 2.1) and a test particle model (Section
2.2), and simulate the flux energy spectrum and production rate of
accelerated electrons by the reconnection electric field in the
solar wind source region (Section 3), to investigate the solar
origin of superhalo electrons.


\section{Numerical Method} 
      \label{S-general}

\subsection{Numerical MHD Model} 
  \label{S-text}
The numerical MHD model used here has been described in details in
\cite{Yang2013}. This section only  gives the basic features and
specifies parmaters/initial conditions for this study. In order to
sustain conversation laws and correct relationships of quantities
across discontinuities in simulations, the 2.5-D resistive MHD
equations, in the Cartesian coordinates ($x, y, z$) with $y$
directed vertically, are written in a conservational form. By
adopting reference values of the plasma density $\rho_0$ ($=
2\times10^{-10}$ kg m$^{-3}$), length $L_0$ (= 1 Mm), and temperature $T_0$ ($=
10^4$ K), these equations are normalized as follows:
\begin{equation}
 \frac{\partial \rho}{\partial
t}+ \nabla \cdot \rho \mathbf{u} = 0
\end{equation}
\begin{equation}
 \frac{\partial \rho \mathbf{u}}{\partial
t}+ \nabla \cdot \left[\rho \mathbf{u} \mathbf{u} + \mathbf{I} ( p +
\frac{1}{2}\mathbf{B}^2 )-\mathbf{B} \mathbf{B}\right] =\rho
\mathbf{g}
\end{equation}
\begin{equation}
 \frac{\partial e}{\partial
t}+ \nabla \cdot \left[\mathbf{u} (e + p +
\frac{1}{2}\mathbf{B}^2)-(\mathbf{u} \cdot
\mathbf{B})\mathbf{B}\right]=\rho \mathbf{u}\cdot \mathbf{g} +
\nabla \cdot ( \mathbf{B}\times \eta\mathbf{j}) - L_r + \nabla \cdot
\mathbf{q} + H +C_N
\end{equation}
\begin{equation}
\frac{\partial \mathbf{B}}{\partial t}+ \nabla \cdot
(\mathbf{u}\mathbf{B}-\mathbf{B}\mathbf{u}) = \eta\nabla^2
\mathbf{B}
\end{equation}
where
\begin{equation}
e=\frac 12 \rho\mathbf{u}^2+\frac{p}{\gamma-1}+\frac 12
\mathbf{B}^2, \ \ \ \ \mathbf{J}  = \nabla \times \mathbf{B}
\end{equation}
correspond to the total energy density and current density,
respectively. Here, $\rho$ is the mass density; $\mathbf{u}=(v_x,
v_y, v_z)$ is the plasma velocity; $p$ is the thermal pressure;
$\mathbf{B}$ denotes the magnetic field; $\mathbf{g}$ ( $= -g
\mathbf{e}_y $, $g$ = const) is the solar gravitational
acceleration; $\gamma \ (= 5/3)$ is the adiabatic index; $L_r$
represents radiative losses; $\nabla \cdot \mathbf{q}$ gives the
anisotropic thermal conduction; $C_N$ is the Newton cooling term;
$H$ is the parameterized heating term; and $\eta$ is the magnetic
resistivity. Here, we assume two resistivity models of the magnetic
reconnection: anomalous resistivity to trigger single X-line
reconnection, and uniform resistivity to trigger multiple X-line
reconnection.

At the Sun, once  the current density $J$ is larger than the
threshold  of  current-driven micro-instability (such as ion
acoustic instability),  large enough local diffusion can be
triggered to increase the local resistivity $\eta$ by orders of
magnitude (e.g., $10^{7}$), the so-called anomalous resistivity
\citep{Treumann2001, Buchner2005}. Many theoretical analysis and
simulations have reported that the anomalous resistivity strongly
depends on $J$ \citep{Sagdeev1967, Davidson1975, Petkaki2008,
Wu2010, Bai2010}, e.g., in a linear or power-law form. Based on the
MHD simulations by \cite{Sato1979}, \cite{Ugai1992} and
\cite{Otto2001}, here we set $\eta$ in the anomalous resistivity
model to be a function of the current density:
\begin{equation} \label{eq:6}
\eta= \left\{ \begin{aligned}
         0, \hspace{1.7cm}   \ \  J \leq J_c \\
          \eta_0(J-J_c)^\alpha, \ \ \  J  > J_c
                          \end{aligned} \right.
                          \end{equation}
where $\eta_0$ and $\alpha$ are the resistivity parameters, and
$J_c$ is the current-density threshold above which the anomalous
resistivity is triggered. As suggested by \cite{Treumann2001} and
\cite{Buchner2005}, $\eta_0$ and $J_c$ are set to be 0.001 $L_0 V_0$
and 0.5 $B_0 L_0^{-1} \mu^{-1}$, respectively, where $V_0( = \sqrt{R
T_0}$, $R$ is gas constant$)$ is the reference velocity, and $B_0 (=
\sqrt{\mu \rho_0} V_0$, $\mu$ is magnetic permeability$)$ is the
reference magnetic strength. For $\alpha$, we simulate for four
different values, 0.5, 1.0 \citep{Ugai1992}, 2.0 \citep{Otto2001},
and 3.0 \citep{Sato1979}.

At the Sun, on the other hand, the collisional resistivity is
generally too small ($\ll$ $10^{-6}\ \Omega $ m) to be resolved
under current computation ability, and to make dissipation difficult
\citep{Cargill2012}. In the uniform resistivity reconnection model,
thus we simulate with an enhanced $\eta$, for three values of
$8\times 10^{-4}$, $2\times 10^{-4}$ and $5\times 10^{-5}$
\citep{Biskamp1980, Samtaney2009, Onofri2004, Onofri2006}. For a
smaller uniform $\eta$ (e.g., $10^{-5}$), however, the
upward-traveling electrons can not be accelerated to energies above
50 keV in our reconnection model.

In both resistivity models, the simulation region spans $-13$ Mm
$\leq x \leq$ 0 Mm in the horizontal dimension and 0 Mm $\leq y
\leq$ 15 Mm in the vertical dimension. This region is covered by a
nonuniform grid in both dimensions, with a grid spacing of $\delta x
=\delta y= 25$ km for 0 Mm $\leq y < $ 4 Mm, $\delta x =\delta y= 50
$ km for 4 Mm $ \leq y < 6$ Mm, and $\delta x =\delta y= 100$ km for
$y \geq$ 6 Mm. We use the same boundary conditions defined by
\cite{Yang2013}.

For the initial conditions, we assume the plasma in hydrostatic
equilibrium with the temperature of a hyperbolic tangent function,
and derive the plasma pressure and density from the static equation
and ideal gas equation \citep{Yokoyama1996}. The initial plasma
velocity is set to be 0. The initial magnetic field is set to be an
open funnel potential field given by \cite{Hackenberg2000}, plus a
closed loop potential field generated by two infinite straight-line
currents in the $z-$direction at ($x=-10$ Mm, $y=-0.7$ Mm) and ($x=
10$ Mm, $y=-0.7$ Mm). We also assume the presence of a relatively
weak magnetic field component $(B_{z0})$ in the $z-$direction, to
allow the guiding center approximation for the particle orbits in
the reconnection region, and the effective partcile acceleration
\citep{Mori1998, Browning2001, Wood2005, Li2012}.

\subsection{Test Particle Approach}

Provided that the electron gyro-radius (gyro-period) is much smaller
than the model scale length (characteristic time), the guiding
center approximation is employed in the present simulation. The
relativistic equations of electron motion in guiding center
approximation are given as follows \citep{Gordovskyy2010}.

\begin{equation} \label{approx1}
\frac{d\mathbf{r}}{dt}=
     \mathbf{u}_\bot+\frac{(\Upsilon v_{\parallel})}{\Upsilon}\mathbf{b}
\end{equation}

$$
\mathbf{u}_\bot=\mathbf{u}_E +
\frac{m}{q}\frac{(\Upsilon{v_{\parallel}})^2}{\Upsilon{\kappa}^2 B}
[\mathbf{b}\times(\mathbf{b}\cdot\nabla)\mathbf{b})]
+\frac{m}{q}\frac{\mu_B}{\Upsilon{\kappa}^2 B}
[\mathbf{b}\times(\nabla(\kappa B))]
+\frac{m}{q}\frac{(\Upsilon{v_{\parallel}})}{{\kappa}^2 B}
[\mathbf{b}\times(\mathbf{b}\cdot\nabla)\mathbf{u}_E)]
$$
\begin{equation} \label{approx2}
+\frac{m}{q}\frac{(\Upsilon{v_{\parallel}})}{{\kappa}^2 B}
[\mathbf{b}\times(\mathbf{u}_E\cdot\nabla)\mathbf{b})]
+\frac{m}{q}\frac{\Upsilon}{{\kappa}^2 B}
[\mathbf{b}\times(\mathbf{u}_E\cdot\nabla)\mathbf{u}_E)]
\end{equation}

\begin{equation} \label{approx3}
\frac{d(\Upsilon v_{\parallel})}{dt} =
\frac{q}{m}\mathbf{E}\cdot\mathbf{b} -
\frac{\mu_B}{\Upsilon}(\mathbf{b}\cdot\nabla(\kappa B)) + (\Upsilon
v_{\parallel})\mathbf{u}_E\cdot ((\mathbf{b}\cdot\nabla)\mathbf{b})
+ \Upsilon \mathbf{u}_E\cdot((\mathbf{u}_E \cdot\nabla)\mathbf{b})
\end{equation}

\begin{equation} \label{approx4}
\Upsilon = \sqrt{\frac{c^2+(\Upsilon v_{\parallel})^2+2\mu_B
B}{c^2-u^2}}
\end{equation}

In the above equations, $\mathbf{r}$ is the electron position vector
and $\mathbf{b}$($=\mathbf{B}/B$) is the unit vector of magnetic
field $\mathbf{B}$; $m$ and $q$ are the electron mass and charge;
$c$ is the speed of light; $\mathbf{u}_\bot$ is the electron
velocity component perpendicular to $\mathbf{B}$, including
$\mathbf{E} \times \mathbf{B}$ drifts, curvature drifts and gradient
drifts; $v_\parallel$ is the electron velocity component parallel to
$\mathbf{B}$; $\mu_B$ is the particle magnetic moment. $\Upsilon (=
{1}/{\sqrt{1-v^2/c^2}})$ is the relativistic factor, where $v$ is
the electron speed, and the coefficient $\kappa$ equals to
$\sqrt{1-{u_E}^2/c^2}$, where
$\mathbf{u}_E=\mathbf{E}\times\mathbf{b}$.

Since the acceleration time ($\sim$ 0.001 s) of electrons is much
shorter than the characteristic timescale ($\sim$ 20 s) of the RCS
evolution in the solar wind source region, we simulate the
trajectory and velocity of electrons with quasi-static background
field snapshots from our MHD model. We spatially bi-linearly
interpolate these MHD model results to obtain
$\mathbf{u(\mathbf{r})}$, $\mathbf{B(\mathbf{r})}$ and
$\mathbf{j(\mathbf{r})}$, and then calculate
$\mathbf{E(\mathbf{r})}$ via $\mathbf{E} =- \mathbf{u} \times
\mathbf{B} + \eta \mathbf{j} $. The fourth-order Runge-Kutta method
is employed to numerically integrate the motion equations (7)-(9),
where the time step ($\delta$t) is adaptive \citep{Zhang2014}. We
also assume that electrons are initially distributed uniformly in
the transition region, with a Maxwellian velocity distribution of T
$ \sim 10^5$ K and no bulk velocity. The rectangle in Figures 1 and 6 denotes the
injection region for test electrons. In each case of $\alpha = 0.5,
1.0, 2.0, 3.0$ and $\eta = 8\times 10^{-4}, 2\times 10^{-4}, 5\times
10^{-5}$, about $10^6$ electron orbits are calculated.

\section{Simulation Results}

\subsection{Single X-line Reconnection}

For single X-line reconnection driven by anomalous resistivity, the
overall evolution of the MHD simulation is similar to that in our
previous work \citep{Yang2013}. This MHD simulation is not sensitive
to the values of $\alpha$, consistent with previous studies
\citep{Sato1979, Otto2001}, while $\alpha$ influences the spectral
shape of the accelerated electrons in the test-particle simulation.
In this session, we illustrate the detailed simulation results for
$\alpha=1.0$, followed by brief descriptions for the other three
$\alpha$ values.

\begin{figure}    
   \centerline{\includegraphics[width=0.9\textwidth,clip=]{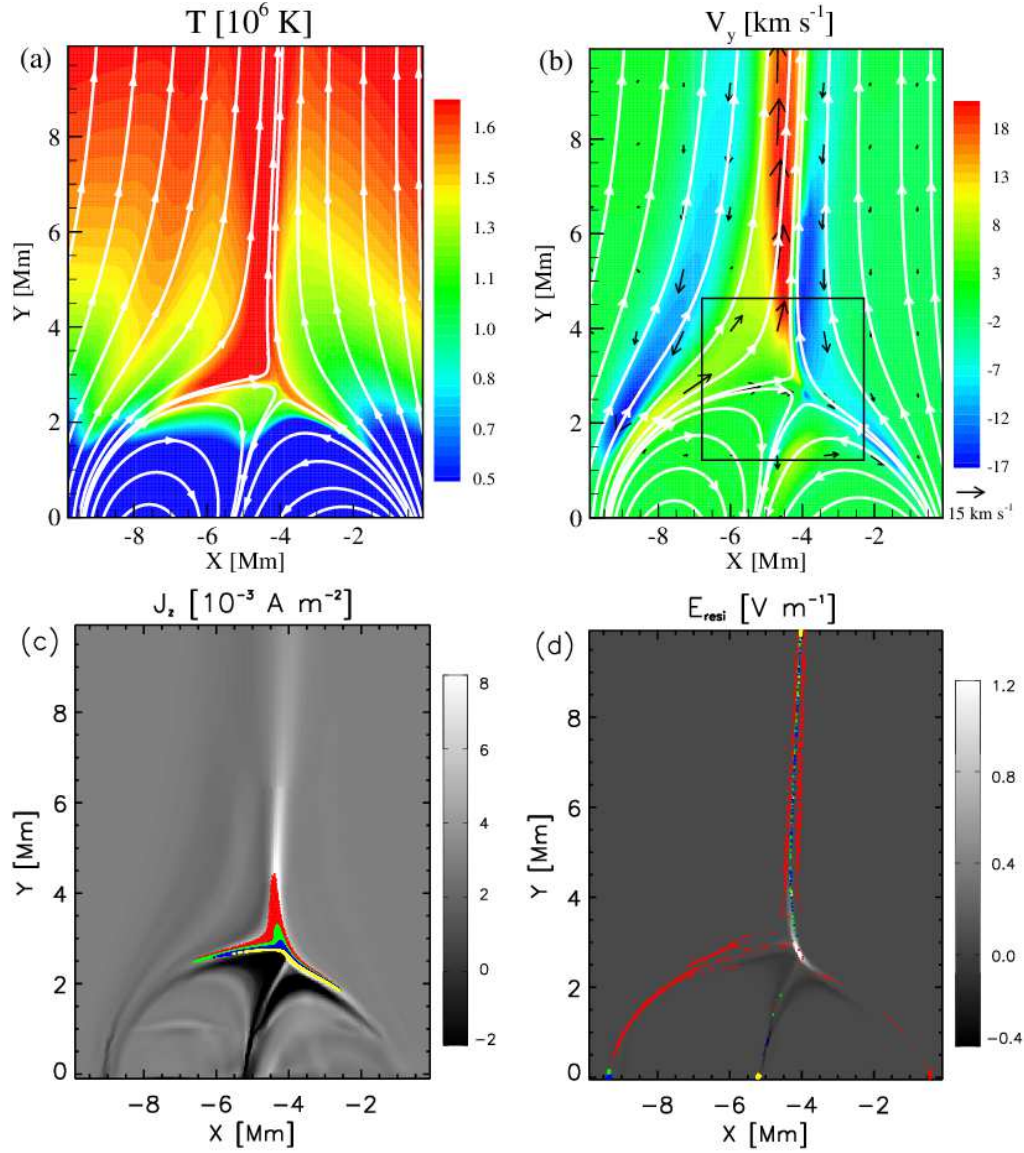}
              }
              \caption{Spatial distributions of the plasma temperature $T$ (a),
vertical velocity $V_y$ (b), out-of-plane current density $J_z$ (c)
and total diffusive electric field $E_{resi}$ (d) at $t=25$ minutes
for the anomalous resistivity parameter $\alpha = 1.0$. In panel a
and b, white streamlines show the magnetic field lines, and black
arrows indicate the plasma velocity. In panel c (d), red, green,
blue and yellow dots denote the initial (last) positions of
electrons over a time interval of 0.1 s, respectively, for the final
energy between 2-5 keV, 6-10 keV, 11-60 keV, and 61-200 keV. This
MHD snapshot is used for the test particle simulation, with the
black rectangle denoting the region where the test electrons are
initially injected.
        }
        \label{Tvy}
   \end{figure}

Figure \ref{Tvy} presents a MHD snapshot for $\alpha = 1.0$, with
the spatial distribution of the temperature $T$ (a), vertical
velocity $V_y$ (b), out-of-plane current density $J_z$ (c) and total
diffusive electric field $E_{resi}$ (d), at $t=25$ minutes (the
early stage of reconnection). We note that carried by the horizontal
flow implemented in the bottom boundary, the hot dense loop is
driven to reconnect with the ambient open coronal field. Such
reconnection produces both upward and downward outflows in the
reconnection region, while the post-reconnection pressure gradient
causes a second upward flow along the newly-opened magnetic field
lines. In the reconnection region, the large current density $J$
forms, and the temperature there is enhanced due to Joule
dissipation. Once $J$ exceeds the threshold $J_c$, the anomalous
resistivity will be switched on. As shown in Figure \ref{Tvy}(d),
therefore, the diffusive electric field $E_{resi}(=\eta J)$ builds
up in the reconnection region, approximately cospatial with the
current sheet.

\begin{figure}
   \centerline{\includegraphics[width=1.\textwidth,clip=]{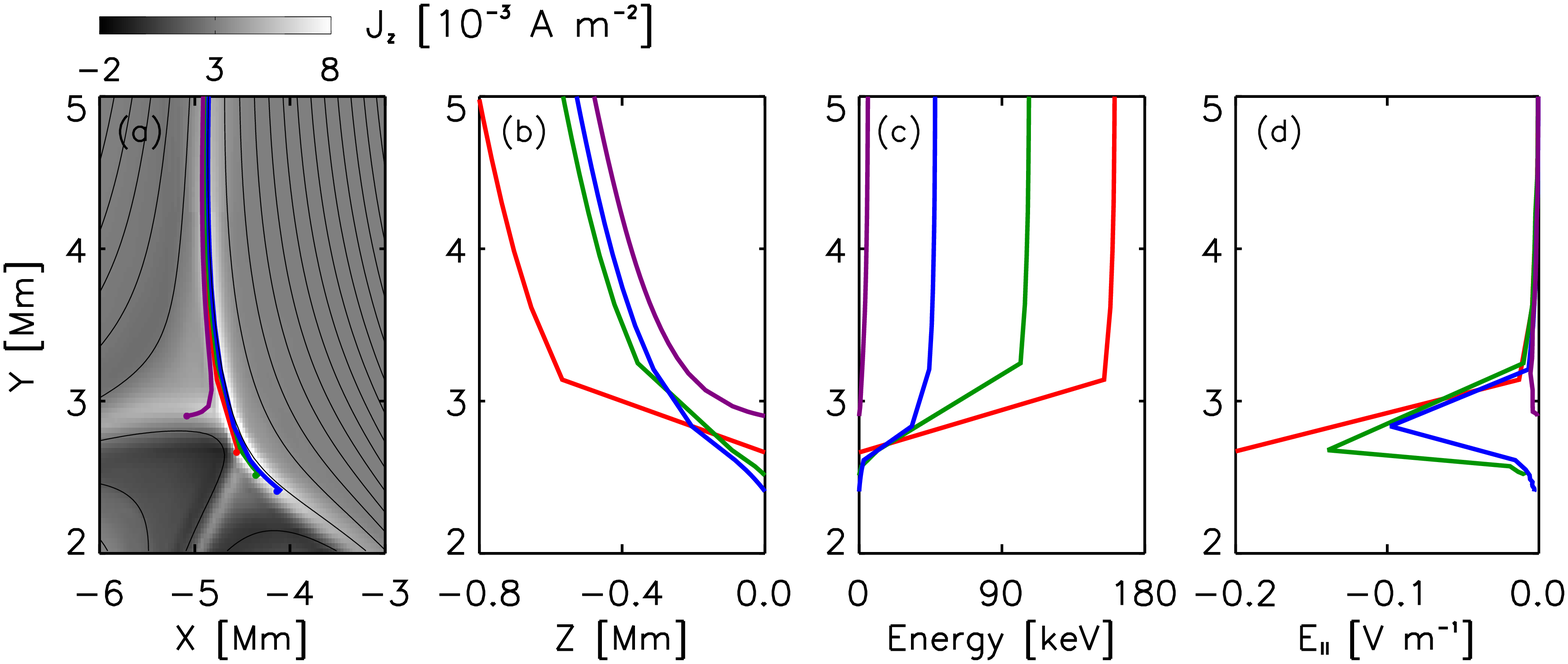}
              }
              \caption{(a): the trajectories in the $x-y$ plane of four sample
electrons with the final energy of 6 keV (purple), 48 keV (blue),
107 keV (green) and 161 keV (red), superimposed in the spatial
distribution of current density $J$, for $\alpha = 1.0$. The colored
dots indicate the electron initial positions. (b): their
trajectories in the $z-y$ plane. (c-d): the electron energy and
$E_\parallel$ versus y, along the trajectories of these four
electrons.
        }
       \label{orbit}
   \end{figure}

Figure \ref{Tvy}(c) and (d) display, respectively, the initial and
last positions of electrons over a time interval of 0.1 s, for the
final energy between 2-5 keV (red), 6-10 keV(green), 11-60 keV
(blue) and 61-200 keV (yellow). In the test particle simulation,
initially about 10$^6$ test electrons are distributed uniformly in a
rectangular region containing the reconnection site (see Figure
\ref{Tvy}(b)), with a Maxwellian velocity distribution of T $\sim 10^5$ K and no bulk velocity.
After a time interval of $\sim$ 0.1 s, about $5\times10^4$ electrons
(5$\%$ of the 10$^6$ test electrons) pass through the reconnection
region, where the diffusive electric field component $E_\parallel$
(parallel to $\mathbf{B}$) is large, and all of them are strongly
accelerated to energies above 2 keV by $E_\parallel$. The closer the
electrons reach to the region with larger $E_\parallel$, the higher
energy they could gain (see Figure \ref{Tvy}(c) and \ref{orbit}).
After the acceleration, these accelerated electrons drift out of the
reconnection region along the magnetic field lines (see Figure
\ref{Tvy}(d)). About half of the accelerated electrons move upwards
along the newly-opened magnetic field lines into IPM, together with
the nascent solar wind flow driven by the reconnection (see Figure
\ref{Tvy}(d)). On the other hand, the other half of accelerated
electrons move downwards into the lower atmosphere, and they would
collide with the ambient dense plasma to emit Hard X-rays via
non-thermal bremsstrahlung. Such Hard X-ray emissions would be very
weak,  probably contributing to the quiet-Sun Hard X-rays
\citep[e.g.,][]{Hannah2010}, since the observed flux of superhalo
electrons at 1 AU is only $\sim 10^5 - 10^6$ of the peak flux of
typical solar energetic electron events associated with Hard X-ray
bursts.

Figure \ref{orbit}(a) and (b) show the trajectories of four sample
electrons with final energy of 6 keV (purple), 48 keV (blue), 107
keV (green) and 161 keV (red). Although the simulation itself is two
dimensional in x and y, the $z$-displacement of electrons is
calculated by integrating $v_z$ over time. Figure \ref{orbit}(c) and
(d) show, respectively, the electron energy and parallel electric
field $E_\parallel$ along these four trajectories. As electrons move
along the magnetic field lines and approach the reconnection region
with non-trivial $E_\parallel$, they start to be energized. The
electrons reaching the very center of the reconnection region can be
accelerated (by large $E_\parallel$) by several orders of magnitude,
within a very short time (e.g., 0.01 s). Meanwhile, electrons move
in the negative z-direction and reach a maximum displacement up to
$\sim$ 0.7 Mm (see Figure \ref{orbit}(b)). Afterwards, electrons
leave the reconnection region and move along magnetic field lines
mainly in the $x-y$ plane; the acceleration dramatically decreases.

\begin{figure}
   \centerline{\includegraphics[width=0.6\textwidth,clip=]{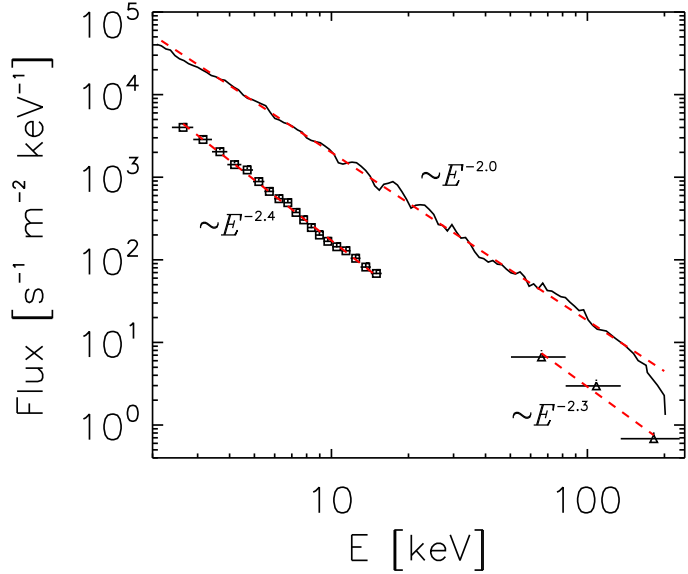}
              }
              \caption{Flux versus energy spectrum of upward-traveling electrons
with the final energy from $\sim$ 2 keV to 200 keV for $\alpha =
1.0$. The squares and asterisks represent the observed superhalo
electrons at 1 AU  from \cite{Wang2012}, with the flux shifted by
five orders of magnitude to compare with the simulation results. The
red-dash straight lines represent a power-law fit to the simulations
results and observations.
        }
       \label{Flux1}
   \end{figure}

Figures \ref{Tvy} and \ref{orbit} suggest that the electric and
magnetic field configurations built up by the magnetic reconnection
in the solar wind source region are capable of accelerating
electrons from thermal to superthermal energies. Assuming a
continuous electron injection into the magnetic reconnection region,
we calculate the flux of upward-traveling electrons around $x = -5 $
Mm and $y = 10 $ Mm. For $\alpha = 1.0$ (see Figure \ref{Flux1}),
the flux versus energy spectrum of electrons at $\sim$ 2-100 keV
fits to a power-law distribution, $f(E) \sim E^{-2.0}$. This
spectral index $\delta$ of 2.0 is consistent with the average index
($2.35\pm0.45$) of superhalo electrons observed in situ during
quiet-time periods \citep{Wang2012}. We also note that in the
present simulation, as the magnetic reconnection evolves, the
spectral shape of accelerated electrons doesn't change very much.

\begin{figure}
   \centerline{\includegraphics[width=0.9\textwidth,clip=]{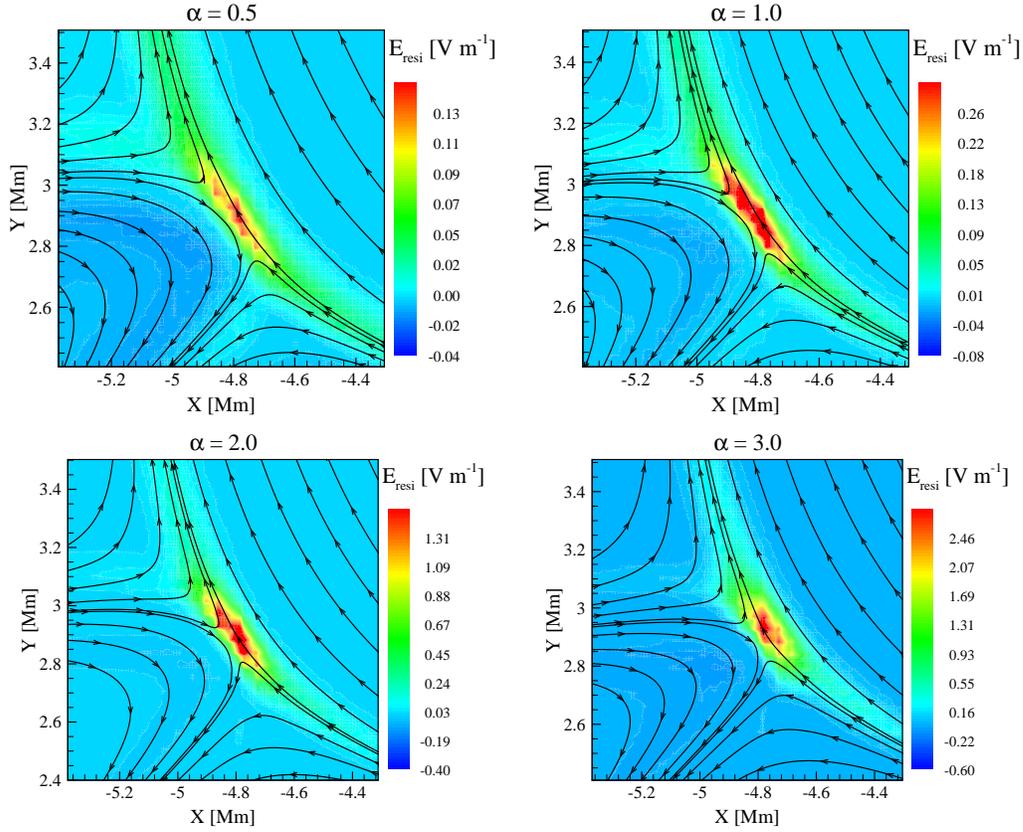}
              }
              \caption{A zoomed-in view of spatial distributions of total diffusive electric field
$E_{resi}$ at $t=25$ minutes for four simulation cases with
$\alpha$=0.5, 1.0, 2.0 and 3.0, in the area around the reconnection
region. Streamlines show the magnetic field lines.
        }
       \label{Figure4}
   \end{figure}

\begin{figure}
   \centerline{\includegraphics[width=0.9\textwidth,clip=]{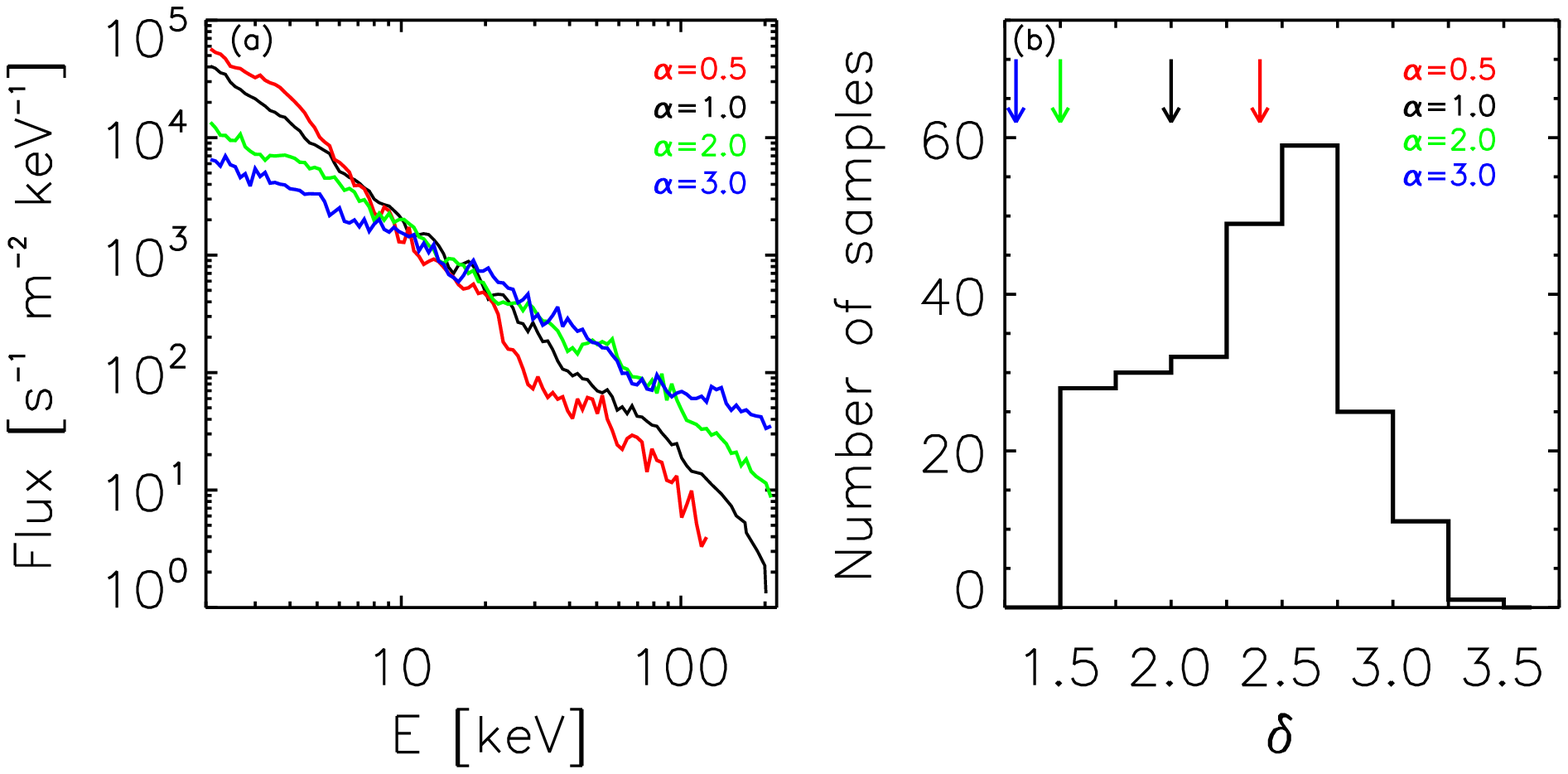}
              }
              \caption{a): Comparison of the flux versus energy spectra of
accelerated electrons between $\alpha$=0.5, 1.0, 2.0 and 3.0, with
the other parameters fixed. (b): Histogram of the observed power-law
index  of superhalo electrons from \cite{Wang2012}, with the colored
arrows indicating the simulated power-law indexes for the four
different $\alpha$.
        }
       \label{Figure5}
   \end{figure}

In the anomalous resistivity model, the energy spectral shape is the
most sensitive to the resistivity parameter $\alpha$. Here, we also
simulate for $\alpha$ = 0.5, 2.0 and 3.0, with the other parameters
fixed. Figure \ref{Figure4} compares the spatial distribution of
total diffusive electric field $E_{resi}$ around the reconnection
region for $\alpha=0.5, 1.0, 2.0$ and 3.0. In the reconnection with
asymmetric inflow, the strong diffusive electric field $E_{resi}$ is
built up on the strong field side of RCS.  As $\alpha$ increases,
the RCS becomes less flat and $E_{resi}$ becomes larger, so more
electrons can be accelerated to higher energies and the resultant
spectrum would become harder (see Figure \ref{Figure5}(a)). This is
consistent with the simulation by \cite{Zharkova2005}. In the four
$\alpha$ cases, the flux energy spectra of accelerated electrons
above 2 keV all exhibit a single power-law function.  When $\alpha$
= 0.5 and 1.0, the simulated spectral index $\delta$ is consistent
with the observations of superhalo electrons, while when $\alpha$
increases to 3.0, $\delta$ decreases to $\sim$ 1.3, harder than the
observations (see Figure \ref{Figure5}(b)).

\subsection{Multiple X-line Reconnection}

\begin{figure}
   \centerline{\includegraphics[width=0.6\textwidth,clip=]{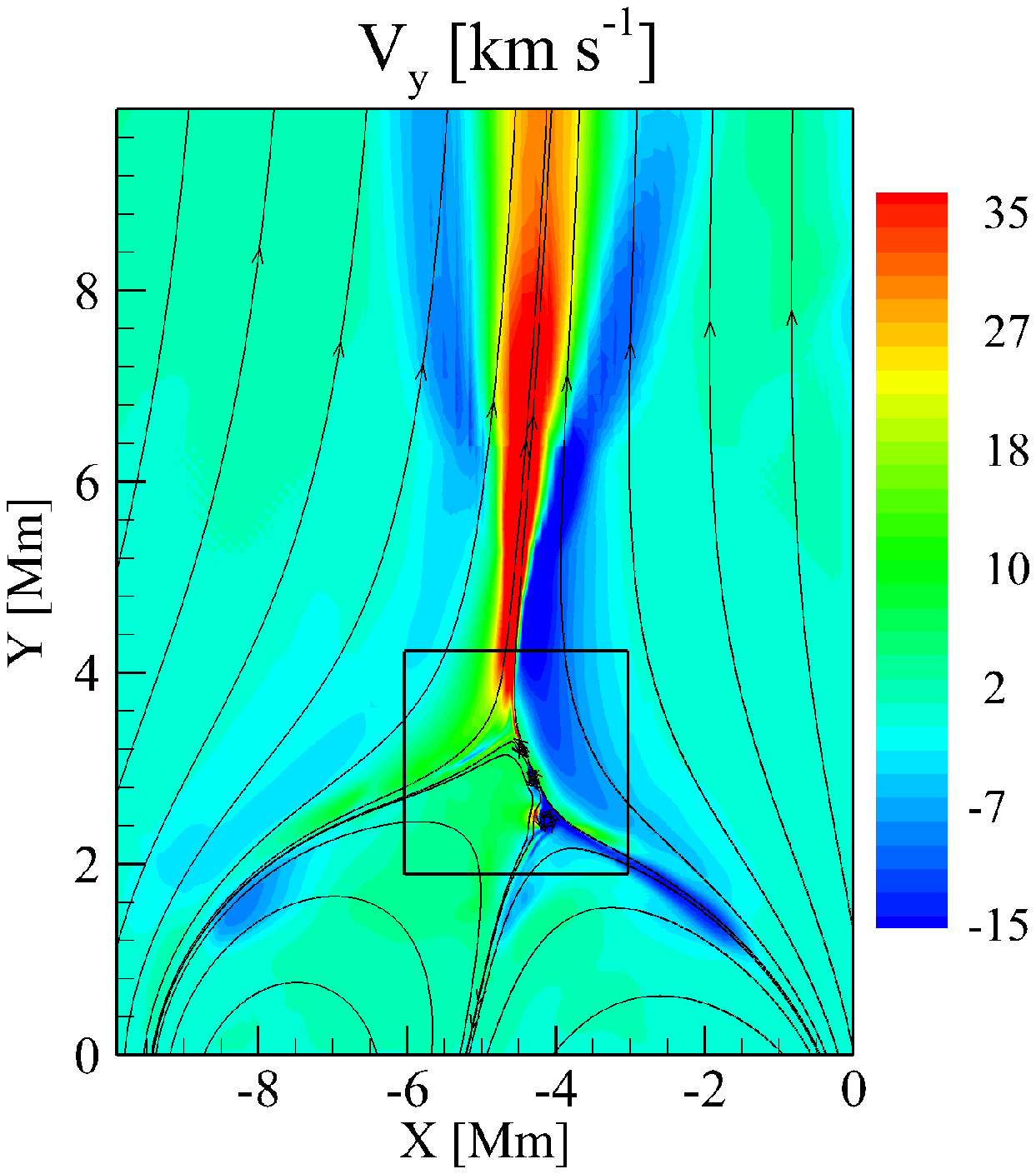}
              }
              \caption{Spatial distributions of the plasma vertical velocity $V_y$
at $t=25$ minutes for uniform resistivity $\eta = 2 \times 10^{-4}$,
with streamlines showing the magnetic field lines. This MHD snapshot
is used for the test particle simulation, with the black rectangle
denoting the region where test electrons are initially injected.
        }
       \label{Figure6}
   \end{figure}

For multiple X-line reconnection driven by small uniform
resistivity, we illustrate the detailed simulation results for $\eta
= 2\times 10^{-4}$, followed by brief descriptions for $\eta$ =
$8\times 10^{-4}$ and $5\times 10^{-5}$. Figure \ref{Figure6} shows
the modeled spatial distributions of the plasma vertical velocity
$V_y$ at $t=25$ minutes (the early stage of reconnection) for $\eta
= 2\times 10^{-4}$. Here the RCS becomes unstable due to tearing
instabilities and is fragmented into several magnetic islands. Such
magnetic reconnection is no longer the standard Sweet-Parker like,
and is inherently time-dependent. In Figure 6, the upward plasma
outflow comes not only from the reconnection region, but also  from
the high-pressure leg of the newly-opened loops, similar to single
X-line reconnection. Therefore, this bursty reconnection would not
change the mass load to the nascent solar wind outflow as stated by
\cite{Yang2013}. Since the acceleration time of electrons is much
smaller than the characteristic timescale of RCS evolution, we can
still use the MHD background field snapshot to conduct test particle
simulation, despite the time-dependent reconnection.

In the test particle simulation, initially about 10$^6$ test
electrons are distributed uniformly in a rectangular region denoted
in Figure \ref{Figure6}, with a Maxwellian velocity distribution of T $\sim
10^5$ K and no bulk velocity. After a time interval of $\sim$ 0.1 s, about $3\times10^4$
electrons have a trajectory through the reconnection region, and all
are accelerated to energies above 2 keV by large $E_\parallel$.
After the acceleration, about half of these high-energy electrons
move upwards along the newly-opened magnetic field lines into IPM.

\begin{figure}
   \centerline{\includegraphics[width=1.\textwidth,clip=]{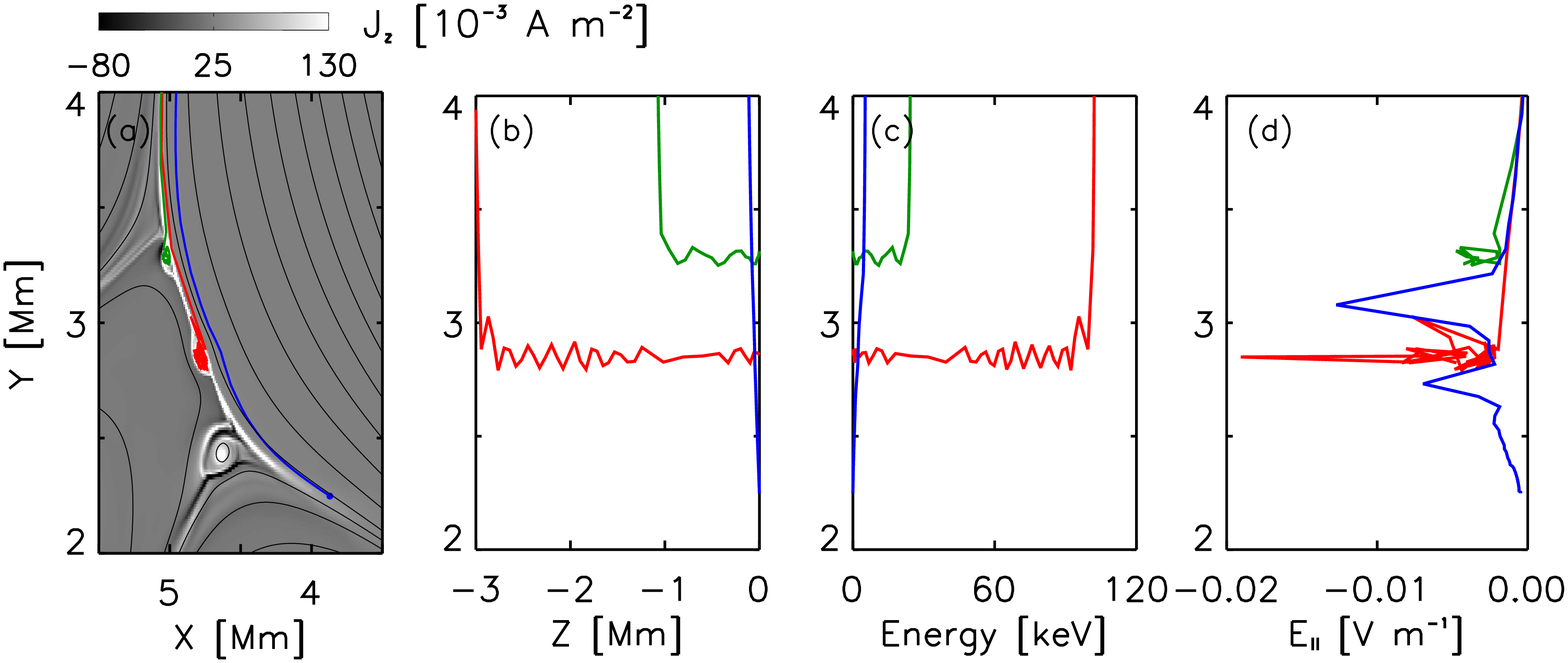}
              }
              \caption{(a): the trajectories in the $x-y$ plane of three sample
accelerated electrons with the final energy of 6 keV (blue), 30 keV
(green) and 110 keV (red), superimposed in the spatial distribution
of current density $J_z$, for uniform resistivity $\eta = 2 \times
10^{-4}$. The colored dots indicate the electron initial positions.
(b): their trajectories in the $z-y$ plane. (c-d): the electron
energy and $E_\parallel$ versus y, along the trajectories of these
three electrons.
        }
       \label{Figure7}
   \end{figure}

Figure \ref{Figure7}(a) and (b) show the trajectories of three
sample accelerated electrons with the final energy of 6 keV (blue),
30 keV (green) and 110 keV (red), respectively, in the x-y and z-y
plane, for $\eta = 2\times 10^{-4}$.  Figure \ref{Figure7}(c) and
(d) display the electron energy and parallel electric field
$E_\parallel$, respectively, along these three trajectories.
Compared with single X-line reconnection, the trajectory and
energization process of electrons become more complicated in
multiple X-line reconnection. For the electrons trapped in magnetic
islands (e.g., see the green and red trajectories), they gain high
energy as they circle around magnetic islands and experience
non-trivial $E_\parallel$. Note that $E_\parallel$ produced in
multiple X-line reconnection is much ( $>$ 10 times) weaker than in
single X-line reconnection, due to small $\eta$. Thus, electrons
would travel a longer distance in the $z$-direction to acquire a
large amount of energy. For the electrons moving freely in open
field lines (e.g., see the blue curve), they can undergo multiple
accelerations by $E_\parallel$ as they pass by multiple X-lines.
However, the final energy gained by these freely moving electrons is
limited due to weak $E_\parallel$.

\begin{figure}
   \centerline{\includegraphics[width=0.9\textwidth,clip=]{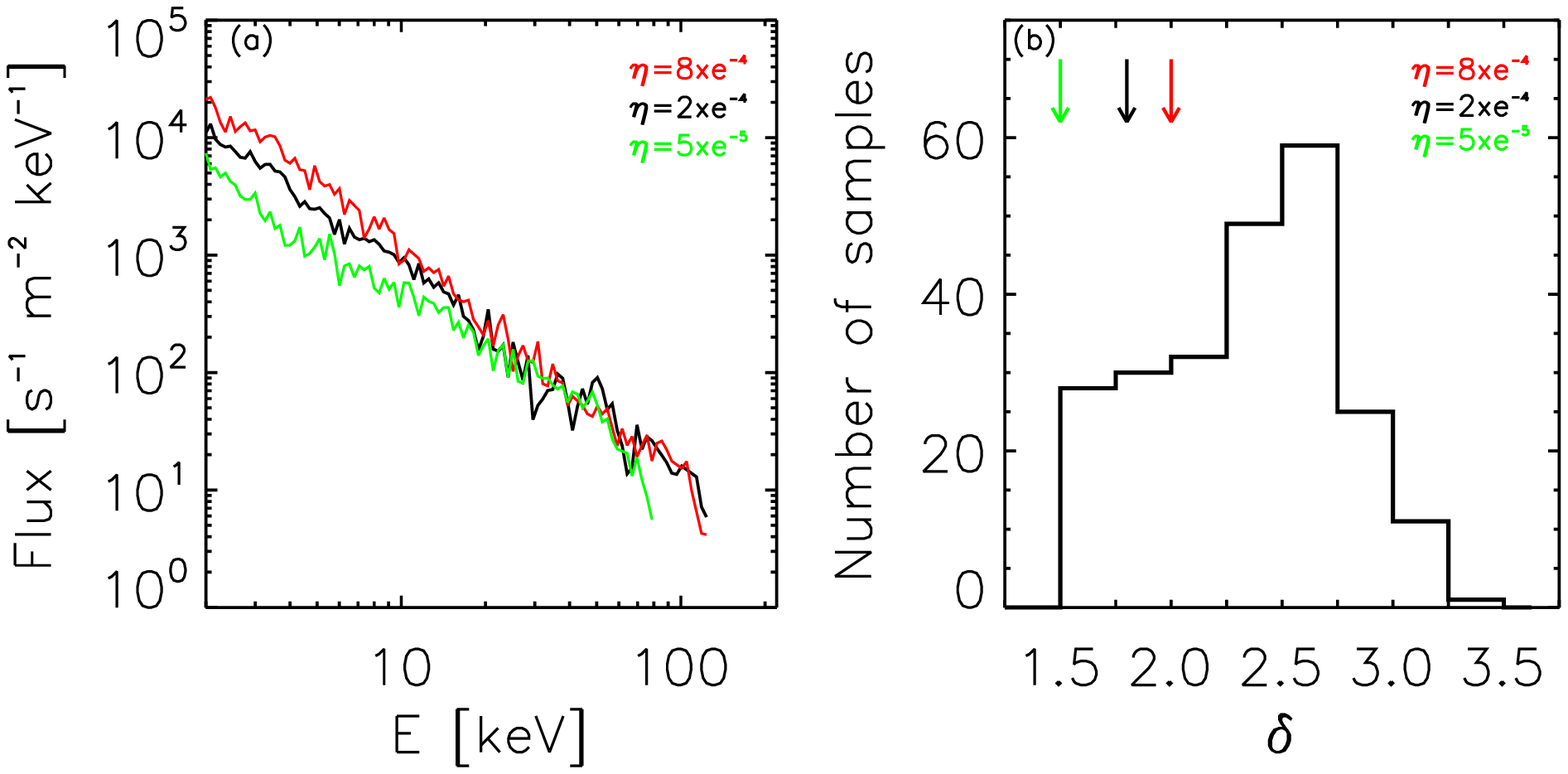}
              }
              \caption{Same as Figure \ref{Figure5} but for uniform resistivity
$\eta$=$8\times 10^{-4}$, $2\times 10^{-4}$, and $5\times 10^{-5}$.
        }
       \label{Figure8}
   \end{figure}

Figure \ref{Figure8} compares the flux versus energy spectra of
accelerated electrons between $\eta=8\times 10^{-4}$, $2\times
10^{-4}$, and $5\times 10^{-5}$, with the other parameters fixed. In
the three cases, the electron energy spectra above $\sim$ 2 keV
generally fit to a single power-law function, with a spectral index
$\delta$ occurring within the range of the observed superhalo
electron indexes during quiet-time periods. As $\eta$ decreases, the
RCS becomes thinner and the $E_{resi}$ away from X-lines gets
smaller, so less electrons are produced at energies below $\sim$ 20
keV (although the higher-energy electrons remain essentially
unchanged) and the resultant spectrum becomes harder. Our
simulations also show that when $\eta$ decreases below
$\sim10^{-5}$, the upward-traveling electrons can not be efficiently
accelerated to energies above 50 keV, and thus the simulated energy
spectrum is no longer a power-law.

\section{Summary and Discussion} 
      \label{S-Conclusion}
In this study, we investigate the generation of superhalo electrons
under the magnetic reconnection model for the solar wind origin.
Using self-consistent electric and magnetic fields obtained from the
MHD reconnection simulation, we conduct test particle simulations to
study the acceleration of electrons in solar wind source region, for
both single X-line reconnection driven by anomalous resistivity and
multiple X-line reconnection driven by small uniform resistivity. We
find that the superhalo electrons may be contributed by the DC
electric field in the magnetic reconnection in the solar wind source
region.

The simulation results show that in both reconnection models,
electrons with an initial Maxwellian velocity distribution of $\sim10^5$ K
can be accelerated to high energies, up to hundreds of keV, mainly
by the DC electric field in the magnetic reconnection. For single
X-line reconnection, electrons gain higher energy as they get closer
to the reconnection region, while for multiple X-line reconnection,
electrons gain high energy as they are trapped in and circle around
magnetic islands. Afterwards, the accelerated electrons follow
magnetic field lines to drift out of the reconnection region. About
half of the accelerated electrons propagate outwards along the
newly-open magnetic field lines into IPM, together with the nascent
solar wind flow driven by the reconnection, while the other half
move downwards into the lower atmosphere. 

In both reconnection models, the energy spectrum of the
upward-traveling electrons generally fits well to a single power-law
at energies of $\sim2-100$ keV. For single X-line reconnection, the
simulated spectral index  $\delta$ features around 2.0, consistent
with the average index ($2.35\pm0.45$) of superhalo electrons
observed during quiet-time periods \citep{Wang2012}. For multiple
X-line reconnection, the simulated $\delta$ is $\sim1.5-2.0$, within
the index range of superhalo electron observations. Among the model
parameters, the resistivity can significantly affect the RCS that,
in turn, changes the energy spectrum of accelerated electrons. For
single X-line reconnection, as the anomalous resistivity parameter
$\alpha$ increases, more electrons are accelerated to high energies,
and thus the simulated spectrum becomes harder. For multiple X-line
reconnection, as uniform resistivity $\eta$ increases, more
electrons can be accelerated to energies below 20 keV and the
simulated spectrum becomes softer.

Since test particle simulation is not self-consistent, we cannot directly estimate the number density of accelerated electrons and thus their density ratio to the solar wind density in the solar wind source region.  Based on the PIC simulation of solar eruptive events \citep{Baumann2012}, here we assume that only $\sim$10\% of the electrons passing through the magnetic reconnection region can be accelerated to energies above 2 keV. Using the physical parameters of the MHD simulation as well as this assumption, the density of upward-traveling accelerated electrons above 2 keV is estimated to be $\sim$10$^{-10}$ of the nascent solar wind flow at y= 10 Mm (see Appendix for details). This is close to the in situ superhalo electron observations \citep{Wang2012}, if the number ratio of superhalo electrons to solar wind plasma doesn’t vary significantly en route to 1 AU. But note that it is unknown whether the 10\% acceleration efficiency assumption is valid in the solar wind source region.

We should note that in the present MHD simulation proposed by
\cite{Yang2013}, only one closed loop is considered to reconnect
with open field lines in the funnel rooted at a chromospheric
network conjunction. After the magnetic flux in this closed loop is
used up, the reconnection process will cease. To form a continuous
and relatively uniform solar wind as observed in the IPM, we need to
consider a large number of independent micro reconnection events
that happen in a limited time period and in many funnels in the wind
source region. Like the reconnection scenario in the solar wind
source region \citep{Tu2005a}, this scenario assumes that the
successive impulsive reconnection events take place at the
interfaces between meso-scale closed loops within the intra-networks
and open funnels rooted from the network junctions, to account for
the continuous presence of superhalo electro population in the IPM.
Those successive impulsive reconnection events may cause local
transient events such as bi-directional plasma jets and network
brightening \citep{Innes1997, Chae2000}, as well as large-scale
quasi-steady Doppler blue shift in the higher transition region and
corona \citep{Tu2005b, He2007, Tian2010}. In the future, we will
develop a 3-D simulation model to describe the formation of
continuous wind from many intermittent micro reconnections and
accordingly develop a more realistic particle acceleration model.

This study is mainly aimed at investigating whether the superhalo
electrons in the IPM could originate from the Sun. The present
model, although simple, is the first to numerically validate this
scenario. In future, we can improve this model by considering a
comprehensive acceleration, instead of only the DC electric field
acceleration. Other mechanisms, such as the turbulence (or
stochastic) acceleration, shock acceleration, or collapsing magnetic
trap acceleration, could also take place in the reconnection region
\citep{Wood2005}. Including these mechanisms would likely allow more
electrons to be accelerated to higher energies, increasing the
density ratio between the accelerated high-energy electrons and
solar wind. Moreover, the Fermi acceleration by reflection from
contracting and merging magnetic islands could occur in the multiple
X-line reconnection \citep{Drake2010}. As a result of the mismatch
between the spatial and temporal scales of the MHD fields and those
of electron motion, our MHD models are unable to address this Fermi
acceleration; instead, the PIC simulation is usually used to study
it \citep{Drake2010, Oka2010}. At the solar wind source
region, however, the very small ion inertial length makes the PIC
simulation subject to resolution constraints, reducing the possible
physical box size that can be simulated to far below the length
scale of the reconnection region. In future, we may also include PIC
simulation to examine the influences of the Fermi acceleration by
contracting and merging magnetic islands on the results presented
here.

Also, we can improve the model by taking into account the effects of
Coulomb collisions at the Sun and the effects of superhalo electrons
propagation in the IPM. Adding the collisions into the acceleration
model would require simulations to be carried out on longer time
scales. In simulations with collisions, \cite{Gordovskyy2013} have
found that the effect of collisions becomes dominant with time,
since the source of acceleration (strong electric field) is
transient and thus gradually disappears, while the source of energy
losses (coulomb collisions) is always present. Therefore, at the
early stage of reconnection, the energy spectra are rather similar
to those obtained in simulations with no collisions, while at the
later stage when electric fields are gradually decaying, the
collisions become dominant. In our test particle simulation for the
early stage of reconnection occurring at transition region, we will
check the effects of Coulomb collisions on the acceleration of
electrons, although such effects may be insignificant. In
addition, during the interplanetary propagation, reflection by the
interplanetary shocks (e.g., CIR shocks), and/or scattering by
wave-particle interaction \citep[e.g.,][]{Yoon2012, Vocks2005}, can
isotropize the angular distribution of superhalo electrons, to form
a nearly isotropic distribution observed at 1 AU. Also note that the evaluated flux and density of the accelerated
electrons in the present model are based upon the value of
acceleration efficiency, which can be sensitive to various MHD model
parameters, especially to the resistivity parameters. In future, we will
consider these respects, to compare the simulations and the
observations in details.


\begin{acknowledgements}
This work at Peking University is supported by NSFC under contract
Nos. 41274172, 41174148, 40890162, 41222032, and 40931055. L.P. is also supported by NSFC under contract Nos. 41304133, 41031066, 41204127, and 41204105, as well as China Postdoctoral Science Foundation. The numerical calculation has been completed on computing system of Peking University.
\end{acknowledgements}

\bibliographystyle{raa}
\bibliography{sola_bibliography_example}

\begin{appendix}
\section{Appendix}
\subsection{Single X-line reconnection}
In the test particle simulation, initially about 10$^6$ test electrons are distributed
uniformly in a rectangular region containing the reconnection site
(see Figure \ref{Tvy}(b)), with a Maxwellian velocity distribution of T $\sim
10^5$ K. After a time interval of $\sim$ 0.1 s, about $5\times10^4$
electrons (5$\%$ of the 10$^6$ test electrons) pass through the
reconnection region, where the diffusive electric field component
$E_\parallel$ (parallel to $\mathbf{B}$) is large, and all of them
are strongly accelerated to energies above 2 keV by $E_\parallel$.
After the acceleration, these accelerated electrons drift out of the
reconnection region along the magnetic field lines (see Figure
\ref{Tvy}(d)), with an average final energy $K_{ave1} \sim 4$ keV
and average final velocity $V_{ave1} \sim 3.5\times10^7$ m s$^{-1}$.
About half of the $5\times10^4$ electrons move upwards along the
newly-opened magnetic field lines into IPM, together with the
nascent solar wind flow driven by the reconnection (see Figure
\ref{Tvy}(d)).

Using the physical parameters of the MHD simulation, we can estimate
the actual flux, density and total energy gain of the accelerated
electrons above 2 keV. In the simulated rectangular region (see
Figure 1(b)) with an area $S_1= 4.5$ Mm $\times$ 3.5 Mm in the $x-y$
plane and a depth $l_1= 1$ cm in the $z-$direction, the actual total
number of initial thermal electrons is
\begin{equation} \label{number1}
 N_1 =  n_0 \times S_1\times l_1 = 1.6\times10^{25},
\end{equation}
where $n_0 = 10^8$ cm$^{-3}$ is the background plasma density.
According to the test particle simulation, about 5$\%$ of the total
electrons would pass through the reconnection region and be all
accelerated within a time interval of $\sim$ 0.1 s. In a self-consistent
simulation, however, not all of them would be accelerated to high energies. The PIC simulation of solar eruptive events by
\cite{Baumann2012} suggests that only $\sim$ 10$\%$ of the electrons
passing through the magnetic reconnection region can be accelerated
to energies above 2 keV. In the present simulation, therefore, the
physical production rate of the $>$ 2 keV electrons can be estimated
as:
\begin{equation} \label{rate1}
P_{e1} = (10\%\times5\%\times N_1)/ 0.1\ \rm{s} = 8\times10^{23}\
\rm{s}^{-1}.
\end{equation}

According to the test-particle simulation, about half of the
accelerated $>$ 2 keV electrons would move upwards along the
newly-opened magnetic field lines into IPM. At $y=10$ Mm, these
electrons cross an area $S_{sup1} = \sim 0.2$ Mm (defined as the
width at the 1/10-fold peak intensity of electron spatial
distribution at the $x-$direction) $\times$ 1 Mm (the 1/10-folding
width in the $z-$direction). Thus, the flux of upward-traveling $>$
2 keV electrons is $f_{sup1} = 0.5\times P_{e1} / S_{sup1}$, and
their number density is
\begin{equation} \label{nd1}
n_{sup1} = f_{sup1} / V_{ave1}  = \sim 5.7 \times 10^{4} \
\rm{m}^{-3},
\end{equation}
At $y = 10$ Mm, the average number density of simulated solar wind
outflow is $n_{sw1} = \sim 2.7 \times 10^{14}\ \rm{m}^{-3}$ from the
MHD simulation. Then the simulated ratio $n_{sup1}/n_{sw1}$ is $~2
\times 10^{-10}$ at this altitude.

\subsection{Multiple X-line reconnection}
In the test particle simulation, initially about 10$^6$ test
electrons are distributed uniformly in a rectangular region denoted
in Figure \ref{Figure6}, with a Maxwellian distribution of T $\sim
10^5$ K. After a time interval of $\sim$ 0.1 s, about $3\times10^4$
electrons (3$\%$ of the 10$^6$ test electrons) pass through the
reconnection region, where the diffusive electric field component
$E_\parallel$ (parallel to $\mathbf{B}$) is large, and all of them
are strongly accelerated to energies above 2 keV by $E_\parallel$.
After the acceleration, these accelerated electrons drift out of the
reconnection region along the magnetic field lines, with an average
final energy $K_{ave2} \sim 3$ keV and average final velocity
$V_{ave2} \sim 3.2\times10^7$ m s$^{-1}$. About half of the
$3\times10^4$ electrons move upwards along the newly-opened magnetic
field lines into IPM, together with the nascent solar wind flow
driven by the reconnection.

Using the physical parameters of the MHD simulation, we can estimate
the actual flux, density and total energy gain of the accelerated
electrons above 2 keV. In the simulated rectangular region (see
Figure \ref{Figure6}) with an area $S_2 = 3$ Mm $\times$ 2 Mm in the
$x-y$ plane and a depth $l_2 = 1$ cm in the z-direction, the actual
total number of initial thermal electrons is
\begin{equation} \label{number2}
 N_2 =  n_0 \times S_2\times l_2 = 6 \times 10^{24},
\end{equation}
where  $n_0 = 10^8$ cm$^{-3}$ is the background plasma density.
According to the test particle simulation, about 3$\%$ of the total
electrons would pass through the reconnection region and be all
accelerated within a time interval of  $\sim$ 0.1 s. In a
self-consistent simulation, however, not all of them would be accelerated to high energies. Then the physical production rate of the
$>$ 2 keV electrons can be estimated as:
\begin{equation} \label{rate2}
P_{e2} = (10\%\times3\%\times N_2)/ 0.1\ \rm{s} = 1.8\times10^{23}\
\rm{s}^{-1},
\end{equation}
for a 10 $\%$ acceleration efficiency in the number of electrons
passing through the reconnection region \citep{Baumann2012}. 

According to the test-particle simulation, about half of the
accelerated $>$ 2 keV electrons would move upwards along the
newly-opened magnetic field lines into IPM. At $y=10$ Mm, these
electrons cross an area $S_{sup2} = \sim 0.4$ Mm (defined as the
width at the 1/10-fold peak intensity of electron spatial
distribution at the $x-$direction, from the test particle
simulation) $\times$ 0.2 Mm (the 1/10-folding width in the
$z-$direction). Thus, the flux of upward-traveling $>$ 2 keV
electrons is $f_{sup2} = 0.5\times P_{e2} / S_{sup2}$, and their
number density is
\begin{equation} \label{nd2}
n_{sup2} = f_{sup2} / V_{ave2}  = \sim 3.4 \times 10^{4} \
\rm{m}^{-3},
\end{equation}
At $y = 10$ Mm, the average number density of simulated solar wind
outflow is $n_{sw2} = \sim 3.4 \times 10^{14}\ \rm{m}^{-3}$ from the
MHD simulation. Then the simulated ratio $n_{sup2}/n_{sw2}$ is
$10^{-10}$ at this altitude.

\end{appendix}

\end{document}